\RequirePackage{fix-cm}
\documentclass[smallextended]{svjour3}       
\smartqed  
\usepackage[round,authoryear]{natbib}
\usepackage{graphicx}
\usepackage{amsmath}
\usepackage{url}
\usepackage{setspace}
\journalname{Computers and Electronics in Agriculture}
\begin{document}

\title{Transfer of Manure as Fertilizer from Livestock Farms to Crop Fields: The Case of Catalonia}


\titlerunning{Transfer of Manure as Fertilizer in Catalonia}        

\author{Andreas Kamilaris \and Andries Engelbrecht  \and Andreas Pitsillides \and Francesc X. Prenafeta-Bold{\'u}}

\authorrunning{Andreas Kamilaris et al.} 

\institute{Andreas Kamilaris \at Research Centre on Interactive Media, Smart Systems and Emerging Technologies (RISE), Nicosia, Cyprus\\
Department of Computer Science, University of Twente, The Netherlands\\
              \email{a.kamilaris@rise.org.cy}	\\
      \and Francesc X. Prenafeta-Bold{\'u} \at Institute of Agriculture and Food Research and Technology (IRTA), Barcelona, Spain\\
              \email{francesc.prenafeta@irta.cat}	\\
           \and Andreas Pitsillides \at Department of Computer Science, University of Cyprus, Nicosia, Cyprus\\
             \email{Andreas.Pitsillides@ucy.ac.cy}    \\
           \and Andries Engelbrecht \at Department of Industrial Engineering, University Of Stellenbosch, South Africa\\
             \email{engel@sun.ac.za}
}

\date{Received: date / Accepted: date}

\maketitle

\begin{abstract}
Intensive livestock production might have a negative environmental impact, by producing large amounts of animal manure, which, if not properly managed, can contaminate nearby water bodies with nutrient excess.
However, if animal manure is exported to nearby crop fields, to be used as organic fertilizer, pollution can be mitigated.
It is a single-objective optimization problem, in regards to finding the best solution for the logistics process of satisfying nutrient needs of crops by means of livestock manure. 
This paper proposes three different approaches to solve the problem: a centralized optimal algorithm (COA), a decentralized nature-inspired cooperative technique, based on the foraging behaviour of ants (AIA), as well as a naive neighbour-based method (NBS), which constitutes the existing practice used today in an ad hoc, uncoordinated manner in Catalonia. 
Results show that the COA approach is 8.5\% more efficient than the AIA. However, the AIA approach is fairer to the farmers and more balanced in terms of average transportation distances that need to be covered by each livestock farmer, while it is 1.07 times more efficient than the NBS.
Our work constitutes the first application of a decentralized AIA to this interesting real-world problem, in a domain where swarm intelligence methods are still under-exploited.

\end{abstract}

\doublespacing

\section{Introduction}
\label{intro}
The central role of the agricultural sector is to provide adequate and high-quality food to an increasing human population,
which is expected to be increased by more than 30\% by 2050 \citep{UNFood}. This means that a significant increase in food production must be achieved.
Because of its importance and relevance, agriculture is a major focus of policy agendas worldwide.
Agriculture is considered as an important contributor to the deterioration of soil, water contamination, as well as air pollution and climate change \citep{bruinsma2003world, vu2007survey}.
Intensive agriculture has been linked to excessive accumulation of soil contaminants \citep{teira2003method},
and significant groundwater pollution with nitrates \citep{stoate2009ecological, garnier1998integrated}.

\begin{figure}
\centering
\includegraphics[width=0.5\linewidth]{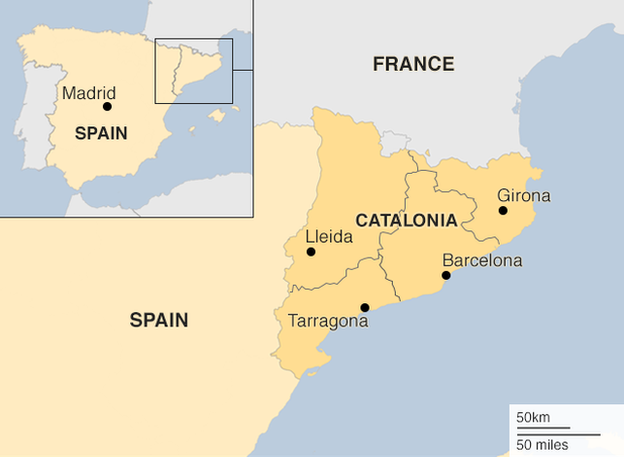}
\caption{Geographical map of Catalonia, Spain.}
\label{fig:Catalonia}
\end{figure}

In particular, intensive livestock farming could have severe negative environmental effects \citep{heinrich2014meat}.
Livestock farms produce large amounts of animal manure, which, if not properly managed, can contaminate nearby underground and aboveground water bodies \citep{cheng2007non, infascelli2010environmental, vu2007survey}.
The autonomous community of Catalonia, located at the north-east part of Spain near the borders with France (see Figure \ref{fig:Catalonia}), is facing this challenge, as livestock farming, mainly swine, has 
contributed to the pollution of the physical environment of the area during the last decades \citep{Kamilaris2017AgriBigCat}.
The high density of livestock in some areas, linked to insufficient accessible arable land, has resulted in severe groundwater pollution with nitrates \citep{directive1991council}. 
Catalonia is one of the European regions with the highest livestock density\footnote{According to the agricultural statistics for 2016, provided by the Ministry of Agriculture, Government of Catalonia.},
with reported numbers of around 7M pigs, 1M cattle and 32M poultry in a geographical area of 32,108 km{$^2$}.

If handled and distributed properly, manure can be applied as organic fertilizer in crop fields that produce different types of fruits and cereals, nuts and vegetables. In this way, the potential contamination of soil and water created by animal manure could be mitigated \citep{he1998preliminary, teira2003method, paudel2009geographic},
while a positive effect on soil acidity and nutrient availability is possible \citep{whalen2000cattle}.
Hence, if the animal manure is efficiently exported at specific seasons of the year to nearby or distant crop fields, manure can eventually become a valuable resource rather than waste \citep{keplinger2006economics, teenstra2014global, oenema2007nutrient}.
To achieve this aim in an optimal manner, the costs of transporting large quantities of manure must be taken into account as a limiting factor in the process of nutrients' transfer from livestock farms to agricultural fields.

This paper proposes two methods to solve the issue of transporting manure from livestock farms to crop fields,
to be used as fertilizer in the territory of Catalonia. The first one is a centralized approach, based on an adapted version of the Dijkstra's algorithm for finding shortest paths together with origin-destination cost matrices \citep{dijkstra1959note}. The second one is a decentralized approach, motivated by the synergistic behaviour of ants at the task of depositing pheromone near food sources, in order to attract more ants to follow their trajectory. This task is foraging, which is achieved by following pheromone trails, and depositing more pheromone on trails during their traversal.
This task creates in a synergistic way promising paths in terms of discovering food \citep{bonabeau1999swarm, garnier2007biological, paredes2017milk}. Intuitively, it can be applied in the context for discovering crop farms in need of fertilizer, similar to the way it has been applied in the past to solve a milk collection problem \citep{paredes2017milk}.

Our contribution in this paper is two-fold: on the one hand, we have solved the problem of transferring animal manure in both centralized and decentralized ways, addressing some limitations of related work (see Section \ref{relWOrk}).
On the other hand, we have proposed and developed a decentralized, nature-inspired technique for a domain (i.e. smart agriculture)
where swarm intelligence methods are still under-exploited, although there is a growing research interest from a computational science perspective \citep{KamilarisSept2018NatComp}.
It is the first attempt to use an ant-inspired algorithm (AIA) for this particular and challenging real-world problem.

The rest of the paper is organized as follows:
Section \ref{relWOrk} describes related work on manure management based on geospatial analysis and on ant-inspired applications in agriculture,
while Section \ref{Methodology} presents our methodology regarding a centralized optimal algorithm (COA), an ant-inspired modelling approach (AIA), as well as a neighbour-based method (NBS). The NBS method constitutes the existing practice used today in an ad hoc, uncoordinated manner in Catalonia \citep{teira1999case, flotats2009manure}.
Section \ref{Results} analyzes the overall findings after applying the proposed methods in the Catalonian context,
and Section \ref{Discussion} discusses the results and comments on the perspectives of this research. Finally, Section \ref{Conclusion} concludes the paper and lists future work.

\section{Related Work}
\label{relWOrk}
Related work involves two main research areas: manure management based on geospatial analysis, facilitated by Geographical Information Systems (GIS) \citep{Kamilaris2018CNNAgri},
as well as applications of ant-inspired techniques in agriculture, facilitated by ant colony optimization (ACO) \citep{dorigo1996ant, dorigo1997ant}. Less relevant work is about network flow solutions applied to other agricultural problems, such as dealing with transportation of live animals to slaughterhouses \citep{oppen2008tabu}, the routing of vehicles for optimized livestock feed distribution \citep{kandiller2017multi} or for biomass transportation \citep{gracia2014application} etc.
Related work in the two main research areas mentioned above is presented below.

\subsection{Transport of Manure for Nutrient Use}
\label{transpManureRelWork}
The idea of transporting surplus manure beyond individual farms for nutrient utilization was proposed in \citep{he1998preliminary},
focusing on animal manure distribution in Michigan.
Teira-Esmatges and Flotats (2003) proposed a methodology to apply manure at a regional and municipal scale in an agronomically correct way,
i.e. by balancing manure distribution to certain crops, based on territorial nitrogen needs and also based on predictions of future needs and availability considering changes in land use.
ValorE \citep{acutis2014valore} is a GIS-based decision support system for livestock manure management,
with a small case study performed at municipality level in the Lombardy region, northern Italy,
indicating the feasibility of manure transfer.

Other researchers proposed approaches to select sites for safe application of animal manure as fertilizer to agricultural land.
Site suitability maps have been created using a GIS-based model in the Netherlands \citep{van1992computer} and in Queensland, Australia \citep{basnet2001selecting}. 
Van Lanen and Wopereis (1992) found that 40\% to 60\% of Dutch rural land was found suitable for slurry injection.
Basnet et al. (2001) presented a method of selecting sites for the safe application of animal waste as fertiliser to agricultural land, concluding that 16\% of the area under study was suitable for animal manure application.

A minimum cost spatial GIS-based model for the transportation of dairy manure was proposed in \citep{paudel2009geographic}.
The model incorporated land use types, locations of dairy farms and farmlands, road networks, and distances
from each dairy farm to receiving farmlands, to identify dairy manure transportation routes that minimize costs relative to environmental and economic constraints.
Finally, an application of ACO to solve the milk blending problem with collection points, determining where the collection points should be located and which milk producers would be allocated to them for delivery is described in \citep{paredes2017milk}.

\subsection{Ant-Inspired Techniques in Agriculture}
Not much research has been done in applying ant-inspired techniques in agriculture.
Few approaches dealing with the application of ACO in agricultural problems have been recorded.
ACO is a probabilistic technique in which artificial ants (i.e. simulation agents) locate optimal solutions by moving through a parameter space representing all possible solutions. ACO generally works by searching for optimal paths in a graph, based on the behaviour of ants seeking a path between their colony and sources of food.
We note that ACO is different than the ant-inspired technique applied to this paper (see Section \ref{AIA}), due to the fact that the agents/ants in our context need to seek multiple paths, in a probabilistic travelling salesman manner. 

Paredes-Belmar et al. (2017) applied ACO to solve the milk blending problem described in the previous section.
Optimal land allocation was investigated in \citep{liu2012multi}, where the ants represented candidate solutions for different types of land use allocation.
Li et al. (2010) used an ACO algorithm for feature selection in a weed recognition problem.
Optimization of field coverage plans for harvesting operations was performed by means of ACO \citep{bakhtiari2013operations}. Finally, ACO was used for
feature selection and classification of hyperspectral remote sensing images \citep{zhou2009feature}, an operation highly relevant to agriculture.

\subsection{Assumptions in Related Work}
The aforementioned related work, presented in Section \ref{transpManureRelWork}. has adopted various assumptions:
\begin{itemize} 
 \item aggregating geographical areas at county-level \citep{he1998preliminary};
 \item selecting generally suitable sites (i.e. crop and pasture areas) to apply animal manure \citep{van1992computer, basnet2001selecting};
 \item not considering transportation distances between livestock and crop farms \citep{he1998preliminary, teira2003method};
 \item not calculating the particular needs of crop fields in nitrogen that depend on the land area and the type of the crop \citep{basnet2001selecting, paudel2009geographic};
 \item not including actual costs involved with the proposed solution \citep{he1998preliminary, paudel2009geographic, teira2003method, basnet2001selecting};
 \item not finding a balanced, fair solution that minimizes the average distance that needs to be covered by the livestock farmers (all aforementioned papers);
 \item approximating the problem by means of only centralized strategies (all aforementioned papers).
 \end{itemize}
 
\section{Problem Modelling and Methods Description}
\label{Methodology}
The overall goal is to solve the problem of how to find an optimal and economic way to distribute animal manure in order to fulfil agricultural fertilization needs.
The purpose of this section is to describe how the problem was modelled using the area of Catalonia as a case study (Section \ref{problemmodel}) and to explain how the objective function was defined (Section \ref{objFunctionDescription}).
Furthermore, this section presents the methods adopted to solve the problem under study. These methods are the centralized optimal algorithm (COA) (Section \ref{COA}), the ant-inspired algorithm (AIA) (Section \ref{AIA}), as well as a method based on neighbour search (NBS) (Section \ref{NBS}). NBS constitutes the prevalent method currently used in the territory \citep{teira1999case, flotats2009manure}, and it has been implemented for comparison purposes.

\subsection{Problem Modelling}
\label{problemmodel}
\begin{figure}[htb]
\begin{center}
\begin{tabular}{l}
	\begin{minipage}{\linewidth}  
		\begin{minipage}{0.50\linewidth}
		\includegraphics[width=\linewidth]
			{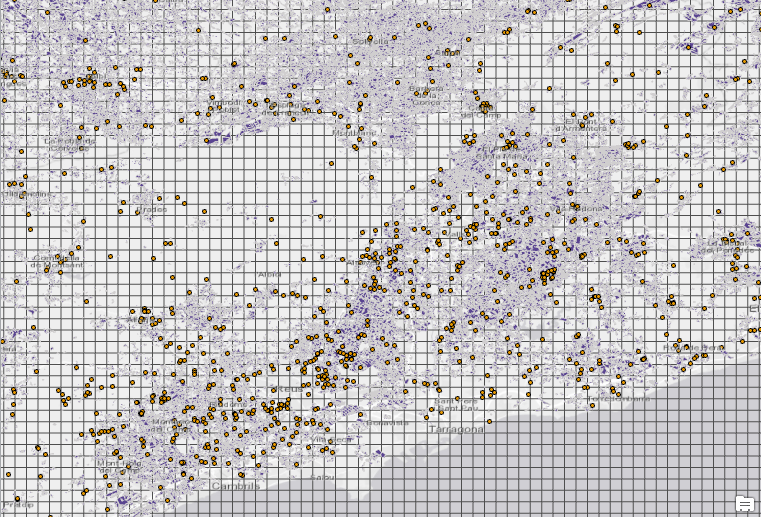}	
		\end{minipage}
		\begin{minipage}{0.49\linewidth}
		\includegraphics[width=\linewidth]
			{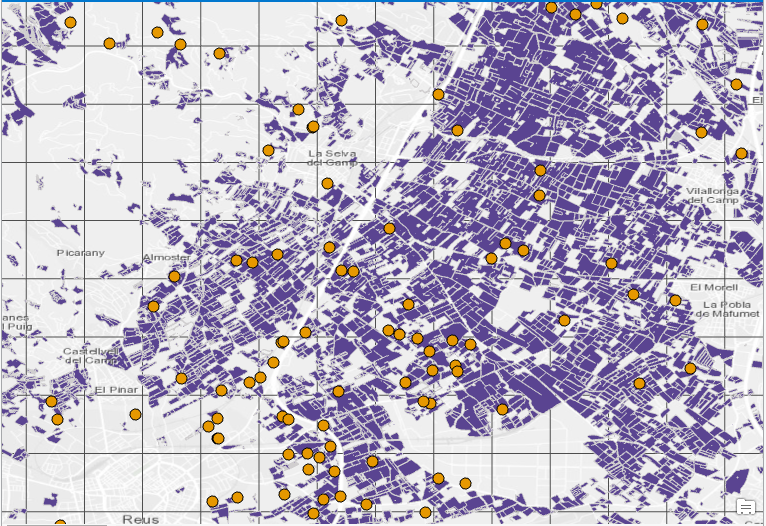}
		\end{minipage}
	\end{minipage}
\end{tabular}
\end{center}
\vspace{-0.3cm}
\caption{Division of the territory of Catalonia in cells of 1 square kilometre each (left). Snapshot is from the area of Cambrils, Reus and Tarragona. Demonstration of livestock/crop farms at grid cells in a dense agricultural area of the region (right). This is a zoom of the map shown on the left. Snapshot is from the area of Reus. Livestock farms are shown as brown circles, and crop fields as blue polygons. The majority of livestock farms raise pigs.}
\label{fig:CataloniaModel}
\end{figure}

To simplify the problem, the geographical area of Catalonia has been divided into a two-dimensional grid, as shown in Figure \ref{fig:CataloniaModel} (left).
In this way, the distances between livestock farms (i.e. original grid cell) and crop fields (e.g. destination grid cell) are easier to compute, considering straight-line grid
cell Manhattan distance as the metric to use; and not actual real distance through the existing transportation network. The centre of the crop field is used for calculations. An approximation to real-world distances is attempted in Section \ref{objFunctionDescription}.

Each crop and livestock farm has been assigned to the grid cell where the farm is physically located, as depicted in Figure \ref{fig:CataloniaModel} (right).
Details about livestock farms (i.e. animal types and census, location etc.) have been provided by the Ministry of Agriculture of Catalonia (Departamento de Agricultura, Ganadería, Pesca y Alimentación, Generalitat de Cataluña) for the year 2016, after signing a confidentiality agreement.
Details about crop fields (i.e. crop type, hectares, irrigation method, location, etc.) have been downloaded from the website of the Ministry\footnote{Ministry of Agriculture of Catalonia. \url{http://agricultura.gencat.cat/ca/serveis/cartografia-sig/aplicatius-tematics-geoinformacio/sigpac/}},
for the year 2015.
For every livestock farm, the yearly amount of manure produced and its equivalent in nitrogen as fertilizer have been
calculated, depending on the type and number of animals on the farm, based on the IPCC guidelines (TIER1) \citep{IPCC2006} and the work in \citep{borhan2012greenhouse}. 
Similarly, for every crop field, the yearly needs in nitrogen have been computed, depending on the crop type and total hectares of land,
according to \citep{RuralCatdossier}.

The estimated total nitrogen needs of crop fields (i.e. 81,960 K-tons of nitrogen) were lower than the availability of nitrogen from animal manure (i.e. 116,746 K-tons of nitrogen). This surplus of nitrogen is evident in Catalonia and has 
contributed to the pollution of the physical environment during the last decades \citep{Kamilaris2017AgriBigCat}.
This means that the produced amount of manure/nitrogen from livestock agriculture has the potential to completely satisfy the total needs of crop farms. This would be particularly important in areas corresponding to the vulnerable zones
defined by the nitrogen EU directive\footnote{The Nitrates Directive of the European Commission. \url{http://ec.europa.eu/environment/water/water-nitrates/index_en.html}}.

Summing up, the total area of Catalonia has been divided into 74,970 grid cells, each representing a $1 \times 1$ square kilometre of physical land. 
Every cell has a unique ID and $(x,y)$ coordinates, ranging between $[1,315]$ for the $x$ coordinate and $[1,238]$ for the $y$ coordinate.
For each grid cell, we are aware of the crop and livestock farms located inside that cell, the manure/nitrogen production (i.e. from the livestock farms) and the needs in nitrogen (i.e. of the crop fields). All types of livestock farms and crop fields have been taken into account.

\subsection{Objective Function}
\label{objFunctionDescription}
The problem under study is a single-objective problem, with the overall goal of optimizing the logistics process of satisfying nutrient needs of crops by means of livestock waste. This goal has the following conflicting sub-objectives:
\begin{enumerate}
 \item The total nitrogen needs at the crop fields have to be satisfied as much as possible.
 \item The total aggregated travel distance covered from the livestock farms to the crop fields, in order to deposit the manure/fertilizer, needs to be as short as possible.
\end{enumerate}

These two sub-objectives can be reformulated as a single one by combining them linearly, assuming the following:
\begin{itemize}
 \item The price of fuel in Catalonia, Spain is 1.27 Euro per liter\footnote{GlobalPetrolPrices. \url{http://es.globalpetrolprices.com/Spain/gasoline_prices/} (for May 2019)}.
 \item The fuel consumption of tanks is 0.203 liters per 100 kilometer \footnote{Natural Resources Canada. \url{http://www.nrcan.gc.ca/energy/efficiency/transportation/cars-light-trucks/buying/16745}}.
 \item Based on the price of fuel in Spain, as given above, the transportation cost per kilometre is 0.257 Euro.
 \item Based on the local monthly average prices for fertilizers in Catalonia\footnote{Ministry of Agriculture of Catalonia. \url{http://agricultura.gencat.cat/ca/departament/dar_estadistiques_observatoris} (ammonium sulphate in May 2019)},
 the value of nitrogen is 0.225 Euro per kilogram.
\end{itemize}

Based on the aforementioned assumptions, the general objective function to be maximized is defined as:
\begin{equation}
\label{combinedObjective}
 GO = (NT \times 0.225 \times l) - (TD \times 0.257 \times g)
\end{equation}

where $NT$ is the total nitrogen transferred in kilograms, and $TD$ is the total distance in kilometres
covered to transport manure, from the livestock to the crop farms. The parameter $l$ aims to capture the nutrient losses of manure during its storage time,
i.e. the time when the manure is stored at the livestock farm until it is transferred to the crop field. Depending on animal type and storage method, nutrient losses vary.
We selected a value of $l=0.60$, which is the average percentage of nitrogen remaining availability in manure according to the animal census of Catalonia,
at an expected storage time of up to three months as solid or liquid manure \citep{rotz2004management}.

Further, the parameter $g$ is a corrective factor aiming to approximate real-world distances, considering that our calculations are based on Manhattan distances between the livestock and the crop farms. The parameter $g$ weights the calculated Manhattan distance by a factor of $g = 1.30$, a value which has been found to be appropriate for approximating real-world distances in semi-rural landscapes \citep{wenzel2017comparing}. 

The objective $GO$ is assumed to be in Euro, as it represents a simplified cost/benefit relationship of the manure transfer problem, i.e. benefit of selling nitrogen to the crop fields and cost of transport needed in order to transfer the nitrogen.
The overall goal is to maximize $GO$, whose value can be translated to gains or losses of each solution of the problem.
$GO$ can take also negative values, which means that some solution would have produced a loss. In this case, the transaction is not executed, since it is not rewarding. For every possible transaction, there is a minimum amount of nitrogen which yields a positive value of the objective function $GO$ (see Table \ref{tab:ParametersAIS}). The simulator compares this minimum amount to the available amount for the transfer and rejects the transfer in case the available content is less than the minimum amount. Thus, for all three methods (COA, AIA and NBS), a transfer is allowed only if the objective $GO$ gives a positive value, based on the current amount of nitrogen and the estimated travel distance, which defines the minimum amount of nitrogen required. Practically, at larger distances, it might not be beneficial to transport manure due to high transportation costs. For example, for a distance of 20 kilometres, there has to be a transfer of at least 51 kilograms of nitrogen for the transfer to be rewarding.

Moreover, there is a hard constraint set by the Ministry of Agriculture, demanding that the maximum distance travelled for manure deposit is $50$ kilometres. The reasoning behind this is that otherwise the travel time required for the transfer would have become significant and should have somehow become included in the calculations. Finally, the Ministry asked to try to maintain the average travel distance and standard deviation from every livestock farm to the crop fields as small as possible, i.e. to keep the proposed solution \textit{well-balanced and fair} for all livestock farms.

\subsection{Centralized Optimal Algorithm}
\label{COA}
A centralized optimal approach has been developed based on the following algorithm, which generalized and adapted the well-known Dijkstra's algorithm for finding shortest paths \citep{cherkassky1996shortest, dijkstra1959note}, together with the use of origin-destination (OD) cost matrices as used in the travelling salesman problem for choosing best routes \citep{lin1973effective}.

\begin{figure}[ht!]
\centering
\vspace{-0.4cm}
\includegraphics[width=1.0\linewidth]{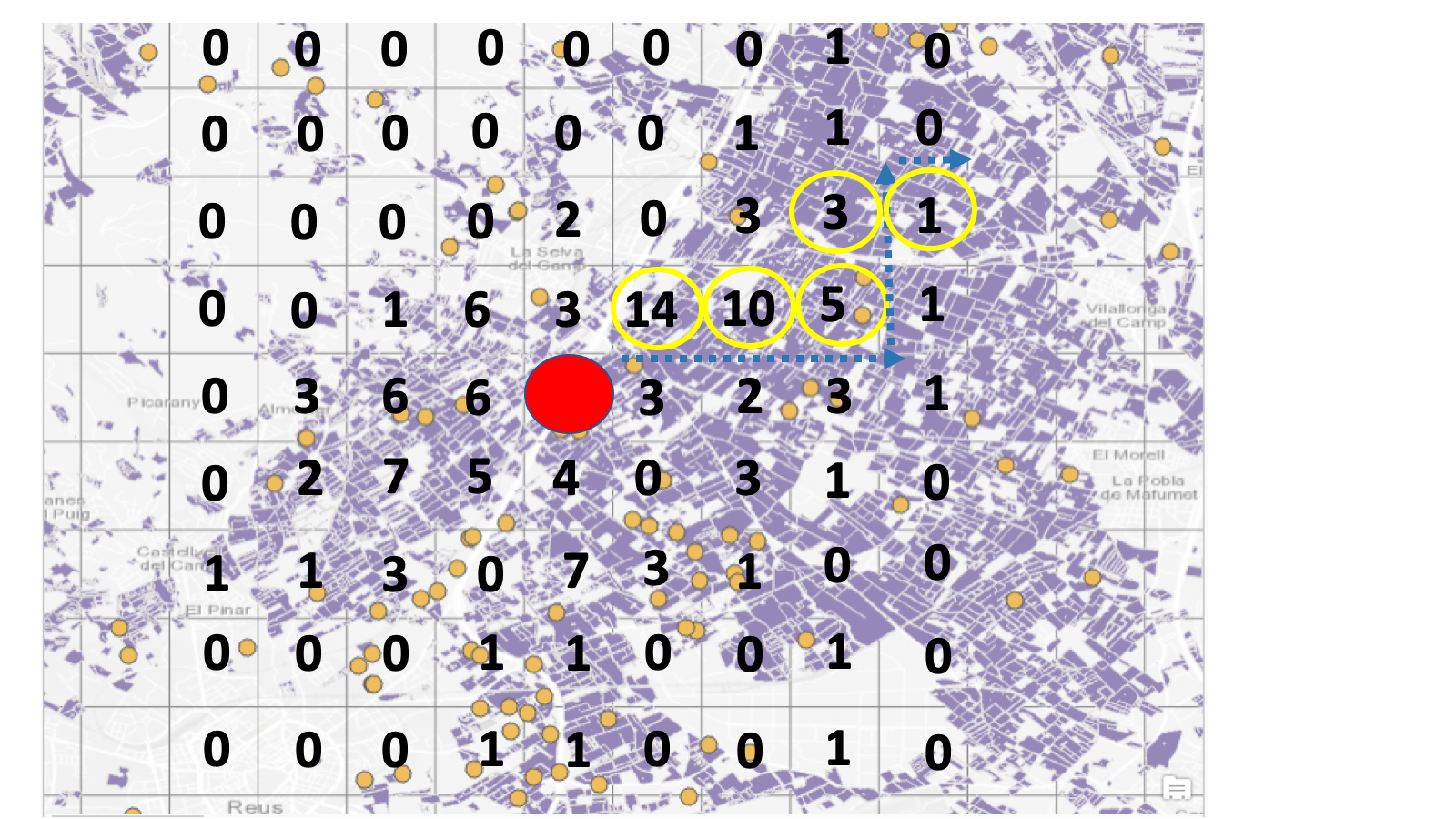}
\caption{Concept of the COA algorithm illustrated.}
\label{fig:COAconcept}
\vspace{-0.1cm}
\end{figure}

Each livestock farm aims to maximize a \textit{local $GO$}, which is the objective function applied only to this farm. In case of conflicts with other livestock farms for common use of resources, the solution that maximizes the \textit{global $GO$}, as defined in Section \ref{combinedObjective}, wins. 

The concept of the algorithm in the context of the problem under study is illustrated in Figure \ref{fig:COAconcept}. Let's assume that the "travelling salesman" is the livestock farm at the red circle. This farm builds its own OD cost matrix, based on the possible values of the local objective function $GO$, applied at each nearby grid cell, up to a Manhattan distance of 50. For reasons of simplicity, Figure \ref{fig:COAconcept} shows the matrix up to a Manhattan distance of 4. We may observe that, generally, grid cells in larger distances have smaller rewards. However, some crop fields located far away might have larger demands in nitrogen, which gives larger values to the local $GO$. It is also possible that crop fields located near competing livestock farms might have reduced demands in nitrogen, as they might have already received nitrogen/fertilizer from these competing farms. After the livestock farm at the red circle builds its OD matrix, then it uses the Dijkstra's algorithm for finding the path that maximizes the local $GO$. In the example of Figure \ref{fig:COAconcept}, this is the path shown by the yellow circles and arrows, which gives a value of $GO=33$. In case of a conflict with another livestock farm (i.e. the two farms share the same grid cell in their paths), the solution maximizing the global objective $GO$ would be considered.

In detail, the algorithm works as follows:

\begin{enumerate}
 \item Every livestock farm makes a complete plan, having visibility of the whole grid in regards to where to transfer manure/nitrogen. The most rewarding paths from the source (i.e. initial position) to all other cells in the grid where crop farms are located, up to a maximum distance of 50 kilometres are calculated, producing an origin-destination cost matrix. The cost or reward of every path is calculated based on the objective function $GO$, considering both the actual transportation distances, and the possible transfer of nitrogen. 
\item Similar to a travelling salesman problem, the possible routes passing from more than one candidate crop farm (i.e. till availability of manure gets satisfied or the hard constrained of 50 kilometres is reached) are added to the origin-destination cost matrix. The goal is to maximize local $GO$, as it applies to the current livestock farm. The selected travel plan involves all the cells that must be visited, starting from the nearest one, which has the highest local $GO$. 
 \item If a conflict appears between the selected travel plans of two livestock farms (i.e. at cell $(x, y)$, where some crop farm is located), the livestock farm involved at the solution that maximizes the global $GO$ wins the conflict. Apparently, if the need of manure/nitrogen at this cell $(x, y)$ is higher than the combined availability of nitrogen by the two livestock farms, then no conflict occurs. 
 \item If the conflict still exists, the livestock farm which has failed in the conflict needs to recalculate a plan that maximizes its local $GO$, this time without considering the cell $(x, y)$ or considering only the remaining need of manure/nitrogen at the crop farm(s) at this cell (i.e. assuming that the livestock farm winning the conflict will deposit its nitrogen there).
 \item Steps 1-4 continue iteratively till there is a global consensus, i.e. no livestock farm can find a better plan to transfer its manure. At the time of a consensus, both the global $GO$ and the individual objective functions for each livestock farm (local $GO$s) have been maximized and cannot be further improved. Any more efforts for conflict resolution do not yield a higher global $GO$.
\end{enumerate}
 
Summing up, the COA solves the problem by the classic Dijkstra's algorithm \cite{dijkstra1959note}, considering a shortest-path problem on an undirected, non-negative, weighted graph. To use the algorithm within the context of the problem under study, the algorithm has been modified to respect the necessary configurations and constraints, i.e. by modelling the weights of the graph to represent both transport distances and crop farms' nitrogen needs, combined using the linear function $GO$. All combinations of visits to nearby farms within 50 kilometres are added to an origin-destination cost matrix, where the most profitable route in terms of maximizing $GO$ is selected. In contrary to the typical travelling salesman problem, here the possible stop locations vary depending on which combinations of candidate crop farms maximize $GO$.

\subsection{Ant-Inspired Algorithm}
\label{AIA}
In general, the synergistic pheromone laying behaviour of ants when discovering food sources
is used as a form of indirect communication, in order to influence the movement of other ants \citep{bonabeau1999swarm, garnier2007biological}.
Pheromone laying was modelled (among others) in the Ant System \citep{dorigo1996ant, dorigo1997ant}, a probabilistic population technique
for combinatorial optimization problems where the search space can be represented by a graph.
The technique exploits the behaviour of ants following links on the graph, constructing paths between their colony and sources of food, to incrementally discover optimal paths, which would form the solution.



In the particular context of the manure transport problem, the foraging behaviour of ants has been adapted  to the problem under study. Each ant (i.e. livestock farm) selects its next position from its current grid position successively
and pseudo-randomly, where the probability of next move depends on the pheromone amounts at the neighbouring grid cells.
At each iteration of the algorithm, each ant is allowed to move at a Manhattan distance of maximum one neighbouring grid cell.
Each ant examines the availability of nitrogen needs by crop fields in its neighbourhood,
and drops pheromone at its current grid cell, proportional to the local needs in nitrogen in order to inform other ants of the demand in manure at nearby crop fields.

In detail, the modelling of the problem according to ant foraging is as follows:
\begin{enumerate}
 \item Every livestock farm simulates an ant.
 \item Every crop field is considered as a potential source of food, analogous to its needs in nitrogen.  At the beginning, the pheromone amount at each grid cell is initialized proportionally to the initial needs in nitrogen by the crop fields physically located inside the grid cell.
 \item Pheromone at each grid cell is updated by pheromone deposits. Ants perform local pheromone updates to the grid cell where they are currently located while moving around, proportional to the amount of food available (i.e. nitrogen needs) in their grid-based neighbourhood of Manhattan distance (i.e. radius) $n$. The pheromone value at each grid cell increases when one or more ants reside at the cell at some point, depositing pheromone, but also evaporates with time.  
 \item Each ant chooses the next link of its path based on information provided by other ants, in the form of pheromone deposits at every grid cell.
 \item Whenever an ant discovers a  crop field with nitrogen needs at its current position (i.e. some grid cell), a transfer of nitrogen is performed from the livestock farm represented by the ant, to the crop field located at that grid cell. In this case, the need for nitrogen at that particular grid cell is reduced accordingly. The manure transaction is recorded by the system as part of the final solution.
 \item If the ant still carries some manure/nitrogen, then it continues to move in the grid up to a maximum Manhattan cell-distance of $m=50$ km from its initial position.
  \item Steps 3-6 continue iteratively till there is a global consensus, i.e. no livestock farm can find a better plan to transfer its manure. At the time of a consensus, the objective function $GO$ has been maximized and cannot be further improved.
\end{enumerate}

The amount of pheromone laid by each ant is calculated based on the amount of existing nitrogen needs at each neighbouring cell within radius $n$. The biological interpretation of $n$ is that it is the distance over which some ant can \textit{sniff} pheromone content released by other ants.
The Manhattan distance calculated is used to penalize neighbours at larger distances, reducing their \textit{contribution} to the pheromone deposits.
The amount of pheromone $\tau_{xy}$, laid by each ant located at grid cell $(x,y)$ at every iteration $t$ of the algorithm, is calculated using:

\begin{equation}
\label{pheromonecreation}
 \tau_{xy}(t) \, = \, \tau_{xy}(t-1) + \sum_{i=x-n}^{x+n} \sum_{j=y-n}^{y+n} NN_{ij} \times \frac{1}{ d_{ijxy}}
\end{equation}
where $\tau_{xy}(t-1)$ is the previous concentration of pheromone at grid cell $(x,y)$,
$NN_{ij}$ represents the food (i.e. needs in nitrogen of the crop field in kilograms) located at grid cell $(i,j)$,
and $d_{ijxy}$ is the Manhattan distance between the ant (i.e. livestock farm) and the food (i.e. crop field). 
The parameter $n$ defines which neighbours at the grid structure would be involved in the calculations of pheromone (i.e. neighbours up to $n$-cell distance).

The probability $p_{kl}$ of an ant to move from grid cell $(x,y)$ to $(k,l)$, is calculated as:
\begin{equation}
\label{antmove}
 p_{kl} \, = \, \frac{\tau_{kl}} {\sum_{i=x-1}^{x+1}\sum_{j=y-1}^{y+1} \tau_{ij} }
\end{equation}

Note that paths with a higher pheromone concentration have higher probability of selection.

At each iteration $t$ of the algorithm, the pheromone concentration $\tau_{xy}(t)$ at every grid cell $(x,y)$ decays/evaporates to promote exploration:
\begin{equation}
\label{pheromoneevap}
 \tau_{xy}(t) \, = \, (1-\varrho) \times \tau_{xy}(t-1)
\end{equation}
where $\varrho$ is the percentage of \textit{pheromone evaporation}.

\subsection{Neighbour-Based Search}
\label{NBS}
For comparison reasons, the method currently used in the Catalonian context was implemented \citep{teira1999case, flotats2009manure}. 
What happens today is that each livestock farmer acts selfishly, trying to find the most appropriate crop field(s) based on the  objective $GO$ (see Section \ref{objFunctionDescription}) to deposit the produced animal manure.

In our implementation, we refer to this method as neighbour-based search (NBS). In reality, the outcome is not optimal, because some farmers might not make the most optimal and rational choice. However,  we have implemented the NBS method assuming the most optimized outcome, as if all farmers made the best possible choice.

The NBS method is described as follows: 
\begin{enumerate}
 \item First, for some cell $(x, y)$, try to transfer nitrogen from the livestock farm to the crop fields located at this same cell (i.e. Manhattan distance 0). Do this for all the livestock farms/grid cells.
 \item Then, if availability of nitrogen still exists, try to transfer nitrogen from the cell $(x, y)$ to the crop fields located at the nearby grid cells $[x \pm 1, y \pm 1]$ (i.e. Manhattan distance 1). Perform this 1-distance calculation for all the livestock farms. 
 \item If the livestock farm cannot find suitable crop farms in the neighbouring cells of Manhattan distance 1, then continue this procedure for grid cells located at increasing distance $k$ each time from cell $(x, y)$.
 At each step $k$, do this $k$-distance calculation for all livestock farms, before moving to a distance $k+1$ (for reasons of fairness).
 \item If some suitable crop farm has been found at distance $k$, then perform the transfer of nitrogen, setting the new position of the livestock farm as the one at the grid cell of distance $k$, where the transfer happened.  Then, move to Step 2. 
 \item If no suitable crop farm has been found at distance $k$, then Steps 3-4 are repeated until either a new crop farm has been found at distance $k+n$  or the availability of nitrogen is completely satisfied, or a maximum distance of $m=50$ (i.e. grid cells distance) has been reached.
 \item Steps 2-5 are repeated for all livestock farms.
\end{enumerate}

\section{Empirical Analysis}
\label{Results}
This section first explains the reasoning towards the tuning of the control parameters of the AIA. Then, it presents and compares the findings obtained by solving the problem of manure transport optimization, using the three methods described in Sections \ref{COA}, \ref{AIA} and \ref{NBS}.

\subsection{AIA Control Parameter Tuning}
\label{AISparameterSetting}
The ant-inspired algorithm introduces the control parameters $n$ and $\varrho$.
Additionally, two more parameters involved in our model are the \textit{maximum cell-distance} $m$ and the \textit{maximum number of iterations}. The former refers to the maximum Manhattan distance
between livestock and crop farms, where nitrogen transfer could be allowed, while the latter defines the maximum number of iterations until the algorithm stops.
The algorithm could stop earlier if no more transfers occur, i.e. all needs are satisfied or no more manure is available.
All parameters involved in the AIA algorithm are listed in Table \ref{tab:ParametersAIS}.

\begin{table}[ht!]
\caption{Control parameters for the AIA algorithm.}
\label{tab:ParametersAIS}       
\begin{tabular}{| p{2.6cm} | p{5.5cm} | p{2.6cm} |}
\hline\noalign{\smallskip}
\bf{Parameter Name} & \bf{Description} & \bf{Value(s)} \\
\noalign{\smallskip}\hline\noalign{\smallskip}
Pheromone evaporation, $\varrho$ & The decay of pheromone deposited by the ants, at each iteration of the algorithm. & 0-100\% \\
\hline
Neighbourhood radius, $n$ & The maximum Manhattan distance, at which neighbouring cells will contribute in calculating pheromone that would be released by the ant. All the cells up to a cell distance $n$ participate in the calculations. & 1-50 grid cells (values up to 65 have been allowed only for testing purposes)\\
\hline
Minimum nitrogen & The minimum amount of nitrogen in kilograms for a transfer to occur, yielding a positive value of the objective $GO$. & 1-150 Kilos, depending on the Manhattan distance between farms. \\
\hline
Maximum cell-distance, $m$ & The maximum Manhattan distance over which transport of animal manure/nitrogen is allowed. & 50 grid cells (values up to 60 have been allowed only for testing purposes)\\
\hline
Maximum iterations & The maximum number of iterations of the AIA algorithm. & 3,000 \\
\noalign{\smallskip}\hline
\end{tabular}
\end{table}

From the parameters listed in Table \ref{tab:ParametersAIS}, the ones whose value needs to be defined are the neighbourhood distance $n$ and the pheromone evaporation coefficient $\varrho$. The former takes values in the range $[0,65]$, ignoring here for reasons of comparison the hard constraint of 50 kilometres, while the latter takes values in the range $[0,100]$.

\begin{figure}
\centering
\includegraphics[width=1.0\linewidth]{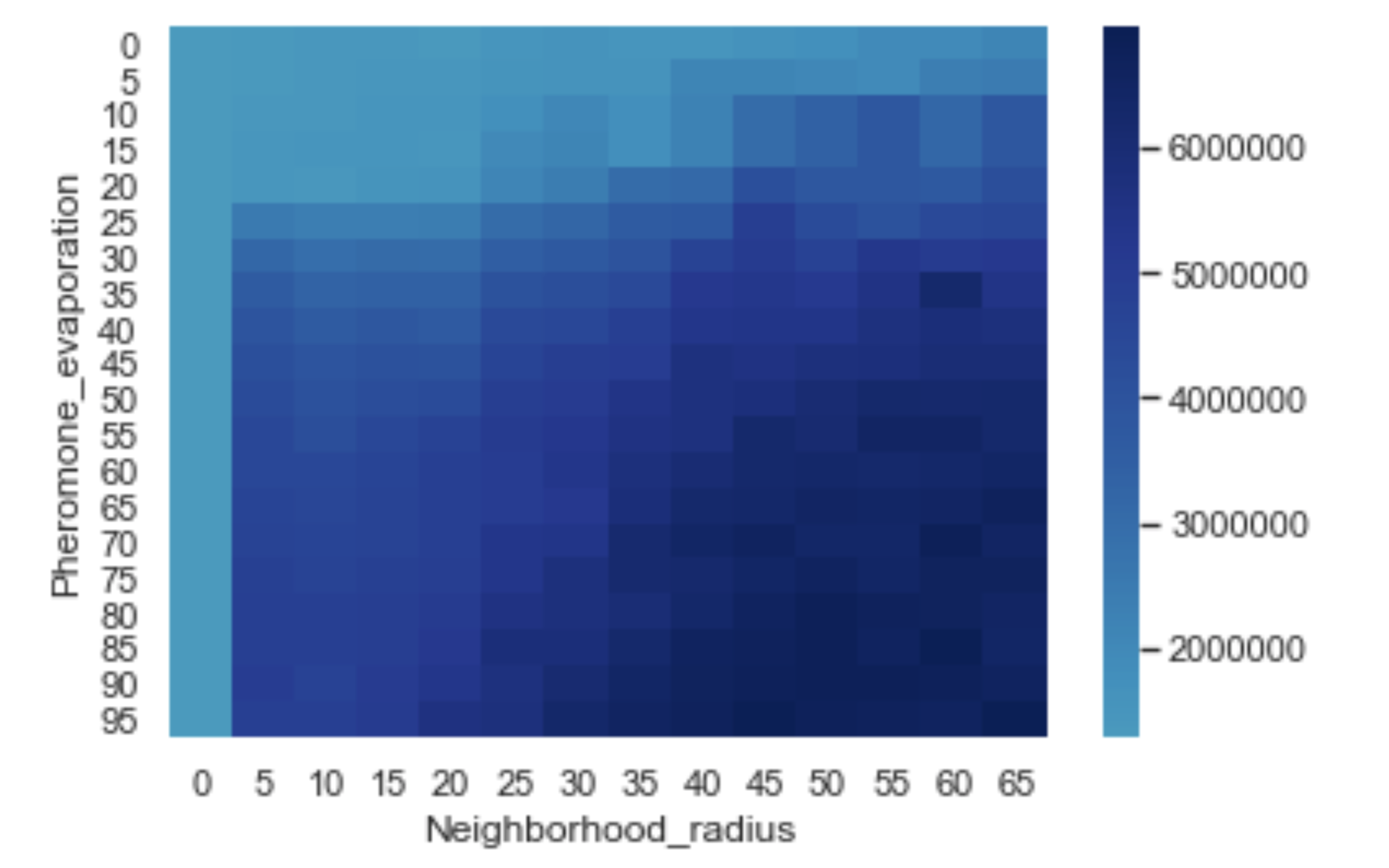}
\caption{Impact of pheromone evaporation $\varrho$ and neighbourhood radius $n$ on the objective $GO$.}
\label{fig:paramsAIA}
\end{figure}

Figure \ref{fig:paramsAIA} depicts the different values of the objective $GO$, at different values of distance $n$ and percentages of $\varrho$. Note that, because the AIA algorithm is stochastic, the results presented below have been averaged over 10 independent runs of the algorithm, with different value pairs of control parameters. The maximum value was recorded for each value pair. Differences between experiments with the same value pairs were very small.

Based on the results presented in Figure \ref{fig:paramsAIA}, a value of pheromone evaporation $\varrho=85\%$ 
and a neighbourhood radius $n=50$ cells-distance
were selected. These parameter values provided a value of $GO = 6,718.069$. We note that values of $n$ larger than the hard constraint of 50 kilometres did not improve $GO$, and have been included for comparisons. We also note that values of $\varrho \in [85,95]$ and $n \in [50,65]$ resulted in very small differences in the $GO$ value.

\subsection{Comparison of COA, AIA and NBS }

\begin{figure}[ht!]
\centering
\includegraphics[width=1.0\linewidth]{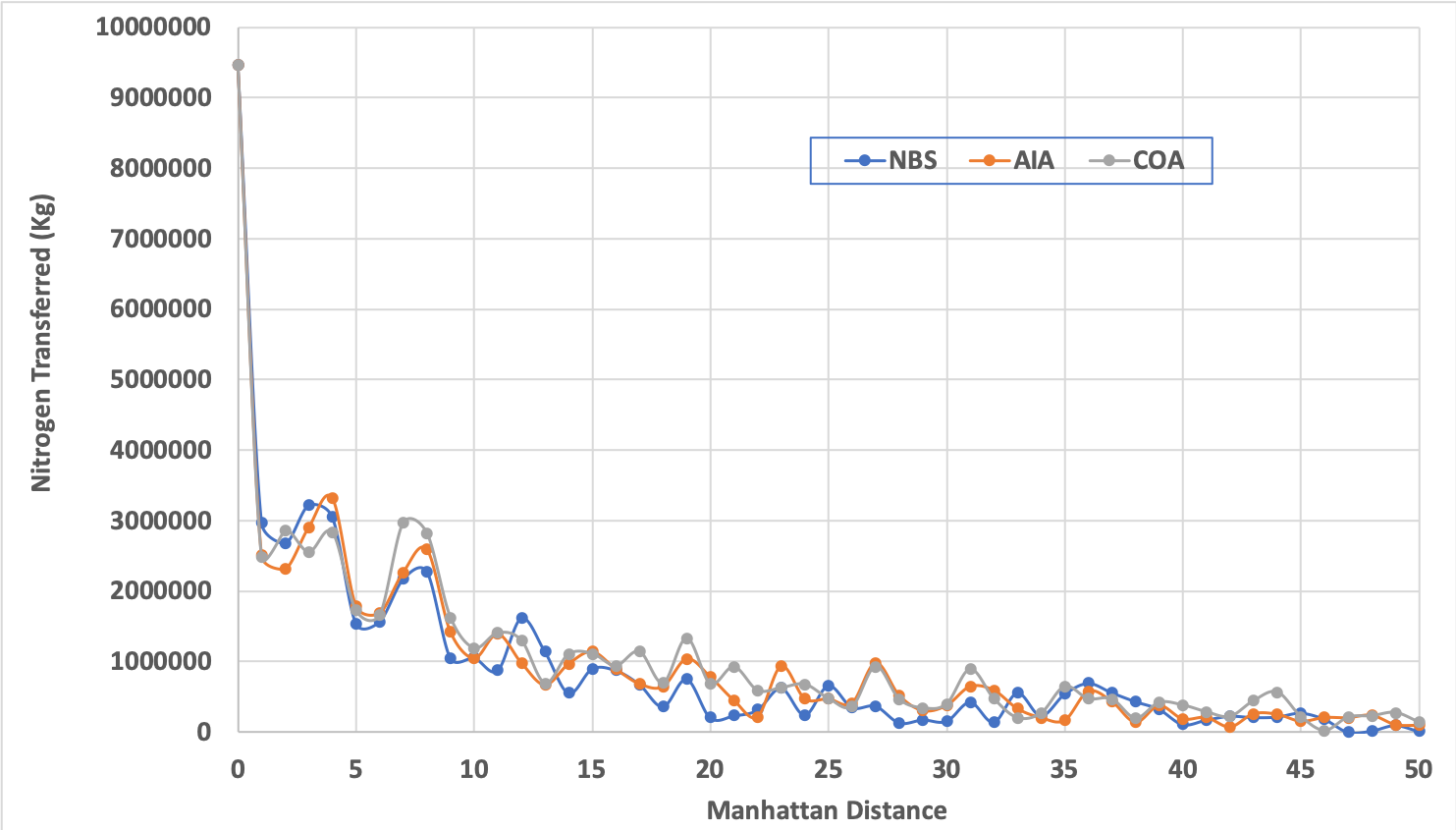}
\caption{A comparison between COA, AIA and NBS for the total nitrogen (in kilos) transferred from livestock to crop farms at different Manhattan distances.}
\label{fig:nitroexchanged}
\vspace{-0.3cm}
\end{figure}

Figure \ref{fig:nitroexchanged} illustrates the total nitrogen transported from livestock to crop farms, for different grid cell Manhattan distances.
COA performs slightly better than AIA, managing to achieve a transfer of 55.3 K-tons of nitrogen (47.4\% from total availability), in comparison to 51,1 K-tons (43.8\% from total availability) for the AIA. NBS transfers less nitrogen than both COA and AIA (47.8 K-tons, 40.9\% from total availability). Hence, in terms of nitrogen transfer, the COA algorithm is 1.08 times more efficient than the AIA algorithm. At the same time, the AIA is 1.07 times more efficient than the NBS.

For all the three approaches, most of the nitrogen transfer happens up to a Manhattan distance of 20 grid cells, after which nitrogen transfer becomes quite low. COA and AIA have larger quantities transferred at lower Manhattan distances (i..e up to 30 grid cells), in comparison to NBS.

\begin{figure}[ht!]
\centering
\includegraphics[width=0.8\linewidth]{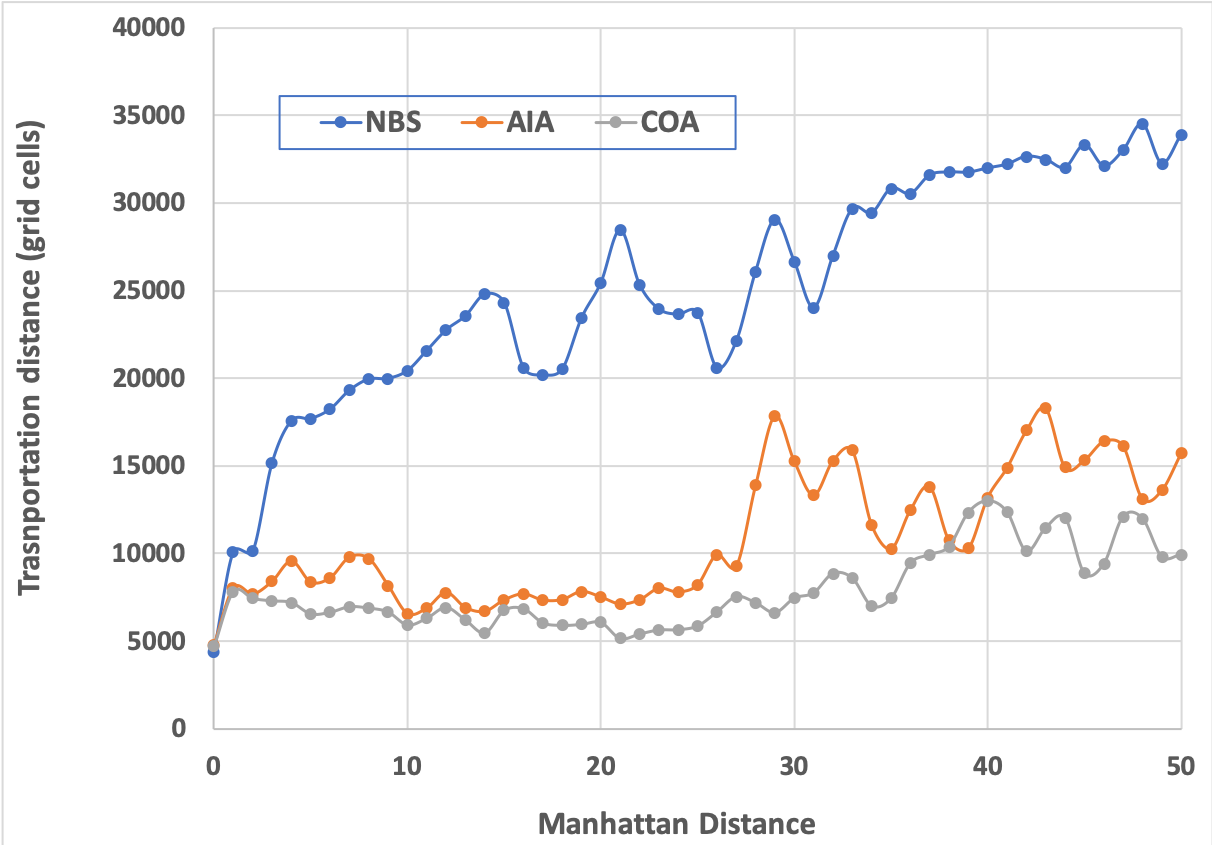}
\caption{Total transportation distance covered between livestock and crop farms, using COA, AIA and NBS at different Manhattan distances.}
\label{fig:transportdistance}
\vspace{-0.1cm}
\end{figure}

Figure \ref{fig:transportdistance} presents the transportation distance covered between livestock and crop farms for every successful transfer of nitrogen, i.e. at each different Manhattan distance recorded for each transfer that took place,
for all the three algorithms. NBS is the least efficient, with a linear increase of transportation distance at larger distances between livestock and crop farms. The COA requires 27\% less distance to be covered than the AIA, while the AIA needs  
57\% less distance than the NBS. Thus, AIA outperforms NBS while COA is more efficient than AIA.

\begin{figure}[ht!]
\centering
\includegraphics[width=1.0\linewidth]{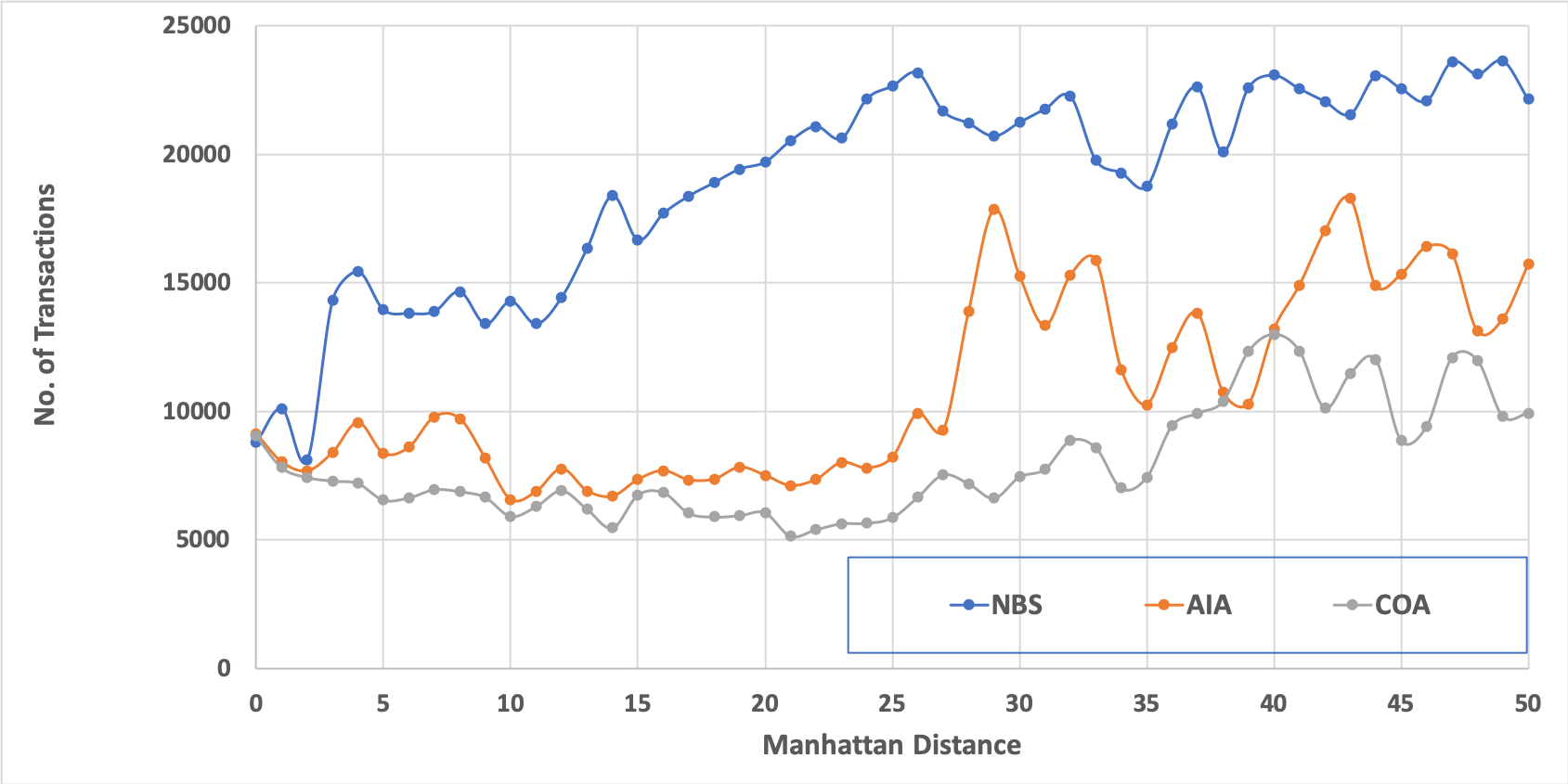}
\caption{Total transactions of animal manure between livestock and crop farms, using COA, AIA and NBS at different Manhattan distances.}
\label{fig:transactions}
\vspace{-0.1cm}
\end{figure}

The total transactions of animal manure performed at different Manhattan distances are presented in Figure \ref{fig:transactions}. The reader can understand the graph in the following way: when there are $x$ transactions for some Manhattan distance $y$, this means that the total transactions that occurred during the simulation, in which the livestock farm involved was located at a Manhattan distance $y$ from the crop field involved, were $x$. COA is the most efficient one, performing less transactions while transferring more manure. AIA performs more transactions than COA in almost all different Manhattan distances, especially 3-8, 27-37 and 41-50. AIA is still much more efficient than NBS. Due to the selfish and competitive behaviour of the livestock farmers at the NBS case, there exist numerous transactions of smaller amounts of animal manure, which cause transactions to increase with distance, especially till a Manhattan distance of 23.

\begin{figure}[ht!]
\centering
\includegraphics[width=0.7\linewidth]{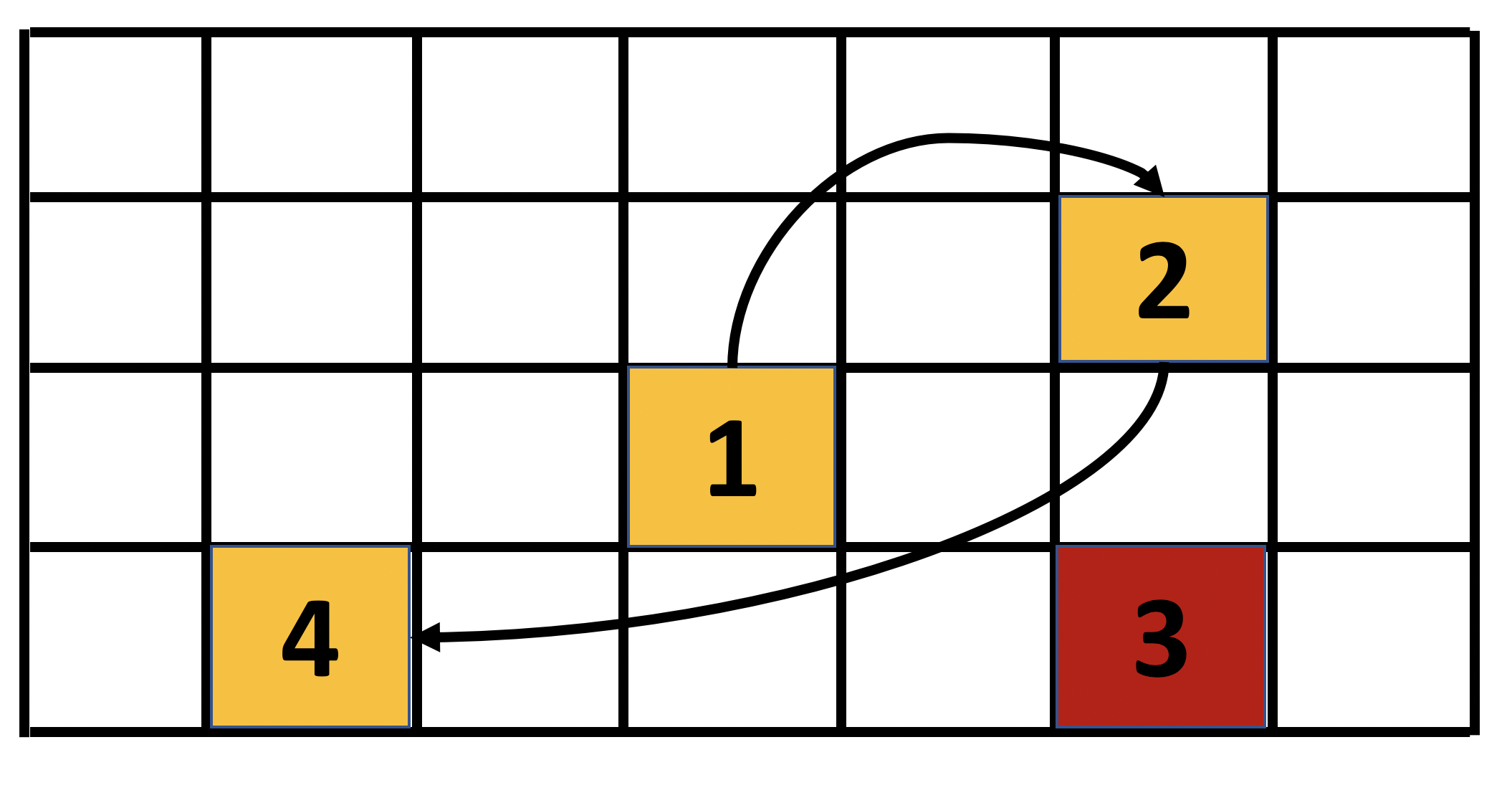}
\caption{An example of not productive behaviour of the NBS method.}
\label{fig:NBSexample}
\vspace{-0.1cm}
\end{figure}

A counter-productive example of the operation of NBS is illustrated in Figure \ref{fig:NBSexample}.
In this example, a livestock farmer physically located at position (1) 
moves nearby east to transfer some manure to position (2), where a crop field is located, knowing that the rest of its available manure would then be placed at position (3).
However, at the next iteration of the algorithm, the need for nitrogen at position (3) becomes satisfied by another rival livestock farmer. Thus, the farmer has to move west from his/her farm's initial position at the next step of the algorithm (i.e. position (4)) in order to deposit the remaining manure/nitrogen. This behaviour increases the overall transportation distance that needs to be covered by the farmer, as indicated in Figure \ref{fig:transportdistance}. The probability of such scenarios is small for the AIA, due to the use of pheromones that coordinate in a more well-balanced way the movement of ants along the Catalonian grid. This probability is zero for COA, because the livestock farms select their strategy a-priori, having complete information of the grid, i.e. based on the distance constraint of 50 kilometres.

Table \ref{tab:summary} summarizes the results of the experiments, including the calculations of the objective $GO$. $GO$ shows that the AIA method is 1.115 times more gainful than the NBS one, however it is 8.5\% less efficient than the COA.
The last two rows of the table denote the average total Manhattan distance that needs to be travelled by each livestock farmer and the standard deviation, in order to perform transfer(s) of animal manure. This average distance is 62 for the COA (with std. deviation of 32), 57 for the AIA method (with std. deviation of 25) and 112 for the NBS (with std. deviation of 78). This relates to the requirement stated in Section \ref{objFunctionDescription}, i.e. the proposed solution must be well-balanced and fair for all livestock farms. The results show that the AIA method is the most well-balanced in terms of transport distance travelled, followed by COA.

\begin{table}[ht!]
\caption{Summarized values of the experiments performed using COA, AIA and NBS.}
\label{tab:summary}       
\centering
\begin{tabular}{| p{5.5cm} | c | c | c | }
\hline\noalign{\smallskip}
\bf{Objective} & \bf{COA} & \bf{AIA} &  \bf{NBS}\\
\noalign{\smallskip}\hline\noalign{\smallskip}
Nitrogen transferred (K-tons)  & 55.385 &  51.124 & 47.786  \\
\hline
Transportation (Manhattan distance)  & 402.379 & 549.829 & 1.276.371  \\
\hline
Objective $GO$ (Euro)  & 7,342.535 &  6,718.069 & 6,024.735  \\
\hline
Average transportation distance of each livestock farm (Manhattan distance)   & 62 & 57 & 112  \\
\hline
Standard deviation of the average transportation distance of each livestock farm (Manhattan distance)  & 32 & 25   & 78 \\
\hline
Running time (minutes)  & 34 &  38 & 31  \\
\noalign{\smallskip}\hline
\end{tabular}
\end{table}

\section{Discussion}
\label{Discussion}
The results indicate that COA is the most efficient solution, outperforming AIA by 8.5\% in reference to a linear objective function $GO$.
This makes sense because COA has complete information of the problem, giving an optimal solution.  However, AIA can be employed to solve the animal manure transport problem in a slightly fairer manner, in terms of balanced transportation distances covered by the livestock farmers. Both COA and AIA solve the problem by reducing effectively the overall transportation distance that needs to be covered from the livestock farms to the crop farm fields, keeping the nitrogen transfer at high percentages.

COA belongs to the class of network flow problems approximated by linear integer programming (ILP). COA runs on a simulator developed by the authors, choosing an adapted generalization of Dijkstra's algorithm for shortest paths, plus the use of origin-destination cost matrices for choosing optimal paths, as used in the travelling salesman problem. The development of a simulator from scratch was decided because of the scale, conditions, objectives and constraints of the problem under study, which made the use of popular ILP solvers (e.g. CPLEX, GLPK, Gurobi) difficult. Besides, the fact that more constraints are expected to be added in the future (see future work in Section \ref{FutureWork} below), influenced the decision to develop a new simulator, for reasons of flexibility and more freedom during future work performed. 

The last row of Table \ref{tab:summary} shows the running time of each algorithm in minutes, on a laptop machine (2,8 GHz Intel Core i7, 6 GB 2133 MHz LPDDR3 RAM). All three algorithms have similar running times, with AIA being the slowest (38 minutes) due to the continuous movement of the ants in the Catalonian virtual grid, till they find a solution or till the constraint of 50 kilometres has been reached. COA has also a considerable running time (34 minutes) because each livestock farm needs to calculate shortest paths to all nearby farms in the radius of 50 kilometres, as well as an origin-destination cost matrix with all possible options. This matrix needs to be created only once, unless conflicts appear (see Section \ref{COA}), in which case some re-calculations need to take place for the livestock farm that has lost the conflict. Due to the fact that not many conflicts have appeared (i.e. less than 400), COA was not much
computationally intensive in the context of the Catalonian area.

The findings indicated that a cut-off Manhattan distance of $50$ was the most appropriate one for the case of Catalonia. 
This cut-off distance is larger than the 30-kilometre cut-off distance selected by \citep{basnet2001selecting} for dairy manure application for the case of Louisiana, USA. A reason for this could be differences in the concentration and topology of the farming industry at the two areas.

\begin{figure}[ht!]
\centering
\includegraphics[width=1.0\linewidth]{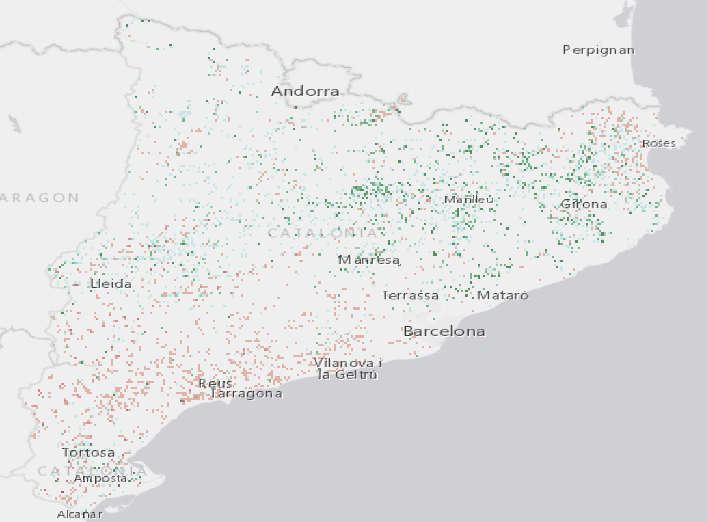}
\caption{The map of Catalonia after the COA has been applied, showing remaining needs in manure (orange color) and remaining availability of manure (green color). The color intensity indicates different needs or availability of manure. For example, darker colours of green and orange correspond to larger availability or needs of manure at some farm. Please note that this map depicts only manure availability and needs of farms after the application of COA. This means that livestock farms whose manure availability is zero and/or crop farms whose needs in manure as fertilizer are zero, do not appear on the map.}
\label{fig:AIAappliedCat}
\vspace{-0.1cm}
\end{figure}

Figure \ref{fig:AIAappliedCat} illustrates how the application of COA in the area of Catalonia affects availability (i.e. green colour) and needs (i.e. orange colour) of manure/nitrogen. We can observe that the algorithm creates separate regions of green- and orange-coloured spots (i.e. livestock and crop farms respectively). The distance between spots of different colour is either larger than 50 kilometres, or there is not enough manure available for the transaction to be gainful, i.e. give positive values to the $GO$ function. Note that darker colours of green and orange correspond to larger availability/needs of manure at some farm respectively. Figure \ref{fig:AIAappliedCat} is another indication that COA solves the problem effectively. A very similar map was produced for the AIA case (although it was  8.5\% less efficient).

As mentioned before, AIA constitutes an important contribution of this paper, due to its decentralized nature.
AIA has potential as an efficient optimization tool in similar problems of a distributed, geospatial nature, and it could well support a dynamic, real-world scenario where supplies in manure and demand in manure/nitrogen could change continuously.
This scenario could be feasible provided that the livestock and crop farmers would be willing to share
information about their animals and their manure and crops respectively. In this case, the AIA algorithm should have been re-designed with faster pheromone evaporations. It is subject of future work.

Moreover, we note that this study constitutes only a demonstration that COA and AIA could be employed for addressing this important problem.
A complete Life-Cycle Analysis (LCA) \citep{curran2008life}, together with Life-Cycle Costing (LCC) \citep{swarr2011environmental}, would consider a more comprehensive coverage of the problem.
For example, 
the profits gained by the algorithms, as summarized in Table \ref{tab:summary}, would be re-considered, taking into account the extra costs needed to maintain the vehicles used for the transfers, i.e. to compensate for the extra kilometres,
as well as the extra time wasted by the livestock farmers or the personnel in charge of realizing the transfers of animal manure. Especially for NBS, having more than triple transport needs than COA as well as double more needs than AIA, this extra cost should be considered as high under a complete LCA.
LCA/LCC could focus on environmental parameters too, incorporating actual costs and comparisons with alternatives. There are environmental consequences by moving large volumes of manure via transportation, not examined in this paper. 

Through this study, we observed that there are considerable differences between larger and smaller livestock farms in terms of the production of animal manure and their overall environmental impact. It would be interesting to compare or enhance our simulator with a hybrid approach/scenario, where larger farms employ local or neighbouring manure processing units and smaller ones participate at this animal manure transfer scheme.  

Finally, it is important to comment that most countries around the world have national policies related to manure management \citep{teenstra2014global}. However, these policies have inconsistencies or they are not well regulated in many countries, especially developing ones \citep{vu2007survey}. Achieving reductions of GHG emissions and meeting renewable energy targets, or lowering the energy costs at farm level are key drivers of manure-related policies, which differ at each country between storage, treatment, digestion, discharge and application \citep{oenema2007nutrient}. A general observation is that manure is not optimally used by farmers generally around the world, especially developing countries \citep{teenstra2014global, vu2007survey, oenema2007nutrient}. Our work aims to contribute to the efforts towards an effective solution to the problem, via application of manure as fertilizer to crop farms, giving insights over the implications of the problem and of its potential solutions.

\subsection{Assumptions and Limitations}
The work in this paper has addressed all the assumptions made in related work (see Section \ref{relWOrk}), being more detailed and complete. Moreover, the AIA solution is completely decentralized, and could be extended for a dynamic scenario (i.e. future work).
However, both the related work and this paper made some additional assumptions, not taking into account the following:
\begin{itemize}
 \item Variation in availability of manure in different periods of the year.
 \item Possibility of larger quantity of manure than the vehicle's capacity to carry, where multiple routes would be needed for the transfer.
 \item Varying crop demands in manure at different seasons. 
 \item Used a simplified objective function to optimize, based on a general estimation of nitrogen value and transport cost (i.e. cost of fuel). Aspects of vehicles' purchase, maintenance and depreciation costs, labour costs etc., have not been considered.
 \item Manure could undergo some \textit{concentration treatment} (e.g. dry cleaning) \citep{teira2003method} in order to reduce the volume transported.
  \item Phosphorous, another fundamental crop nutrient present in manure, has not been considered.
\end{itemize}

An additional important assumption was the modelling via grid cells and Manhattan distances instead of actual, real-world distances. This assumption was considered due to the overall computational complexity of the problem. We tried to mitigate this issue by approximating real-world distances using the corrective factor $g$ in the objective function (see Section \ref{objFunctionDescription}), but this is only a simplified approximation. The factors of faster vs quicker routes, quality of the roads, obstacles such as mountains and city centres requiring additional kilometers to travel, slope of each route, traffic in rush hours, speed limit in different roads, constraints in the routes that trucks are allowed to take, etc. have not been taken into account. Transportation distance relates also to time waste, which has also not been considered. These, together with the assumptions mentioned before, are important aspects of future work, which is discussed below.

\subsection{Future Work}
\label{FutureWork}
Future work will continue to explore the application of the COA and the AIA to this problem, addressing the assumptions made in this paper.
More realistic transportation distances and travel times among farms for manure transport would be considered, as well as dynamic changes in production and need for nitrogen.
This will include the possibility of various routes during the year to transfer manure, calculating more precisely the seasonal effect on the nitrogen content available in the manure which is being reduced through time.
Also, the seasonal differences of various crops will be studied, which might make some crop fields unavailable for manure application at some periods of the year.
Moreover, the costs of the trucks involved in the transport (i.e. purchase, maintenance, depreciation, etc.) will be considered in the objective function, although this is complicated topic due to the possible subsidies that might be provided by the government in order to implement such a manure transport scheme.
Finally, we plan to investigate the use of local or neighbouring manure processing units in selected livestock farms. The complete environmental consequences of the problem under study would be considered too, including the pollution produced by the transportation of manure between farms.

\section{Conclusion}
\label{Conclusion}
This paper addressed the problem of the surplus of animal manure from livestock agriculture, which creates important environmental problems. The paper investigated and suggested a sustainable approach based on nutrient redistribution, where manure was transported as fertilizer from livestock farms to crop fields. Two approaches have been developed: a centralized approach (COA) based on an adapted version of Dijkstra's algorithm for finding shortest paths; as well as a
decentralized one inspired by ant foraging behaviour (AIA). AIA addressed the problem by modelling livestock farms as ants and crop fields as sources of food for the ants.

A comparison between the (centralized) COA approach and the cooperative and decentralized AIA algorithm showed that the COA was 8.5\% more efficient, based on a single-objective function. Both COA and AIA outperformed significantly a (individualist) Neighbour-Based Search (NBS) approach, which resembles the existing practice used today for transport of manure in the region of Catalonia, Spain. The AIA approach was fairer for the farmers and more balanced in terms of average transportation distances that need to be covered by each livestock farmer to transport manure.

Our work constitutes a new application of ant-inspired algorithms to an interesting real-world problem, in a domain where swarm intelligence methods are still under-exploited.

\begin{acknowledgements}
Special thanks to Mr. Jaume Boixadera Llobet and Mr. Mario Carrillo Salagre from the Ministry of Agriculture, Government of Catalonia.
Their feedback, help and advice has been very important in terms of understanding the problem of livestock agriculture in Catalonia and seeking together ways to reduce it.

This research has been supported by the P-SPHERE project, which has received
funding from the European Union's Horizon 2020 research and innovation programme under the Marie Skodowska-Curie grant agreement No 665919,
and also by the CERCA Programme/Generalitat de Catalunya.

Andreas Kamilaris has received funding from the European Union's Horizon 2020 research and innovation programme under grant agreement No 739578 complemented by the Government of the Republic of Cyprus through the Directorate General for European Programmes, Coordination and Development.

Francesc X. Prenafeta-Bold{\'u} belongs to the Consolidated Research Group TERRA (2017 SGR 1290), funded by the Generalitat de Catalunya.
\end{acknowledgements}

\bibliographystyle{plainnat} 
\bibliography{acobib}   

\end{document}